\documentclass[reprint, amsmath,amssymb, aps, prl,superscriptaddress]{revtex4-2}

\usepackage{graphicx}
\usepackage[dvipsnames]{xcolor}
\usepackage{amsmath}
\usepackage{amssymb}
\usepackage{hyperref}
\usepackage[percent]{overpic}
\usepackage{comment}
\usepackage{subcaption}

\usepackage{braket}

\newcommand\eq[1]{Eq.~(\ref{eq:#1})}

\newcommand\fig[1]{Fig.\ref{fig:#1}}

\newcommand\Fig[1]{Figure~\ref{fig:#1}}

\newcommand{\negspace}{\!}

\newcommand{\rsub}[2]{{#1} \negspace {\protect\vphantom{#1}}_{#2}}

\newcommand{\ketsub}[2]{\rsub {\ket{#1}} {#2}}

\usepackage{ulem}

\begin{document}

\graphicspath{{img/}}

\title{Deep reinforcement learning for near-deterministic preparation of cubic- and quartic-phase gates in photonic quantum computing}

\author{Amanuel Anteneh}
\affiliation{440 West Farmington Road, Virginia Beach, VA 23454, USA}
\author{L\'eandre Brunel}
\altaffiliation{Now at Welinq.}
\author{Carlos Gonz\'alez-Arciniegas}
\altaffiliation{Now at Xanadu.}
\affiliation{Department of Physics, University of Virginia, 382 McCormick Rd, Charlottesville, VA 22903, USA}
\author{Olivier Pfister}
\affiliation{Department of Physics, University of Virginia, 382 McCormick Rd, Charlottesville, VA 22903, USA}
\affiliation{Charles L. Brown Department of Electrical and Computer Engineering, University of Virginia, 351 McCormick Road, Charlottesville, VA 22903, USA}
\email{olivier.pfister@gmail.com}

\date{\today}

\begin{abstract}
Cubic-phase states are a sufficient resource for universal quantum computing over continuous variables. We present results from numerical experiments in which deep neural networks are trained via reinforcement learning to control a quantum optical circuit for generating cubic-phase states, with an average success rate of $96\%$.  The only non-Gaussian resource required is photon-number-resolving measurements. We also show that the exact same resources enable the direct generation of a quartic-phase gate, with no need for a cubic gate decomposition.
\end{abstract}


\maketitle

\section{Introduction}
Continuous-variable quantum computing (CVQC) utilizes bosonic field (qumode) encoding in lieu of native qubits. Qumode encoding still enables hybrid bosonic qubit encoding (for example via GKP states~\cite{Gottesman2001}) and also offers the exceptional scalability of quantum optics, be it in free space~\cite{Pysher2011,Chen2014,Yokoyama2013,Yoshikawa2016,Asavanant2019,Larsen2019,Roh2025} or, more recently, on chip~\cite{Yang2021,Jahanbozorgi2023,Wang2024,Jia2025}. As Lloyd and Braunstein pointed out, CVQC is universal if one has access to at-least cubic Hamiltonian evolution in the quantum fields~\cite{Lloyd1999}. It is also important to note that, besides its aforementioned outstanding scalability, CVQC can be made fault tolerant~\cite{Menicucci2014ft}. Last but not least, CVQC offers a unique platform for the quantum simulation of quantum field theory~\cite{Marshall2015a,Briceno2024}. 

An example of cubic Hamiltonian evolution is the cubic-phase gate $\exp(i\gamma Q^3)$, where $\gamma$ is a real parameter and $Q=(a+a^\dag)/\sqrt2$ is the field's amplitude quadrature, $a$ being the photon annihilation operator. Cubic-phase gates have a non-Gaussian Wigner function since Gaussian gates are generated by Hamiltonians at most quadratic in the fields~\cite{Hudson}. When combined with other cubic and quadratic gates, the cubic-phase gate can then yield evolution at any order of the fields, by virtue of the Baker-Campbell-Hausdorff formula~\cite{Lloyd1999}. Gate decompositions in terms of cubic-phase gates have been studied in detail~\cite{Sefi2011,Kalajdzievski2019,Budinger2024}: realizing a quartic gate $\exp(i\delta Q^4)$, for example, requires 29 gates, 15 of which cubic~\cite{Kalajdzievski2019}. In this paper, we show that quartic-phase gates can be generated directly, with no need for a cubic decomposition.  

In the context of measurement-based quantum computing (MBQC), which is the foundation of modern photonic QC~\cite{Menicucci2006,Furusawa2011,Pfister2019,Bartolucci2021,Bourassa2021,Renault2025}, cubic-phase gates can be applied to an arbitrary state $\ket\psi$, to Gaussian corrections left, by teleporting $\ket\psi$ through a cubic-phase state~\cite{Gottesman2001}, 
\begin{align}\label{eq:idealcubic}
\ket\gamma=\sqrt{2\pi}\,e^{i\gamma Q^3}\ketsub0p=\int_{-\infty}^{+\infty} ds\,e^{i\gamma s^3}\ketsub sq,
\end{align}
where $\ketsub s{p,q}$ are the respective eigenstates, of eigenvalue $s\in\mathbb R$, of the phase quadrature [$P=i(a^\dag-a)/\sqrt2$] and amplitude quadrature operators. The cubic-phase state $\ket\gamma$ is therefore a key resource for universal CVQC. %and is analogous to a CV magic state~\cite{Bravyi2005,Baragiola2019}. 

Thinking ``classically,'' one may seek to prepare cubic-phase states by generating cubic waves from a cubic field Hamiltonian. This deterministic generation of cubic-phase states can be done by squeezing a non-Gaussian trisqueezed state, easily obtained in the transmon-qubit cavity microwave fields that experience very large nonlinearities~\cite{Zheng2021_Giulia}. That direct approach is difficult to translate to the optical domain due to the weakness of third-order optical nonlinearities. This could be circumvented in the temporally and spatially confined, guided ultrafast optical regime, where nonlinearities are more efficient due to the higher peak field amplitudes~\cite{Yanagimoto2020}. 

Thinking more ``quantum mechanically,'' one may forgo using a third-order nonlinearity and instead seek to generate nonlinear waves by leveraging their corpuscular nature, e.g.\ using photon-number-resolving (PNR) measurements. This was the original---though probabilistic---proposal of Gottesman, Kitaev, and Preskill (GKP),  employing displaced PNR detection on half of a two-mode-squeezed state~\cite{Gottesman2001}. It was shown in a later study that squeezing on the order of 17 dB and probabilistically detected photon numbers on the order of 50 would be needed to reach QC-useful values of $\gamma$~\cite{Ghose2007}. 
While PNR measurements have since been demonstrated at the 100-photon level~\cite{Eaton2023}, this remains a demanding specification. Other, non-probabilistic methods were proposed to reach smaller values of $\gamma$~\cite{Marshall2015,Marek2018}. 

In this paper, we generalize the GKP protocol to an iterative process with an added in-loop displacement. When driven by a deep neural network trained by reinforcement learning, the process is near-deterministic, reaching a 96\% success rate for the generation of a cubic-phase state with $\gamma=0.2$, using squeezing no higher than 10 dB, low displacements, and much lower PNR measurement values. We then give an even simpler protocol for directly, though probabilistically, generating a quartic-phase state. We expect this protocol to form the basis for another, similar machine learning (ML) algorithm. 

\section{Reinforcement learning for quantum state preparation}

\subsection{Quantum optical circuit}

The quantum optical circuit we employed is displayed in \fig a. 
\begin{figure}[ht]
\vglue -.05in
\begin{center}
\includegraphics[width=\columnwidth]{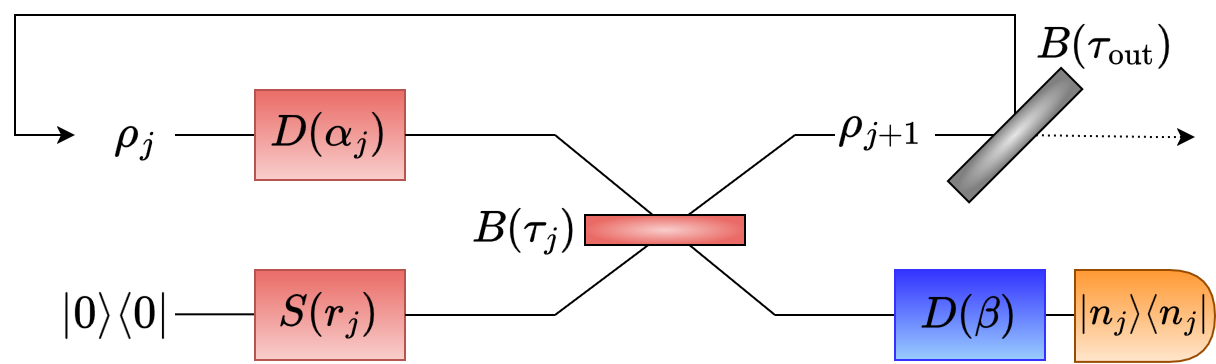}   %\fbox
\end{center}
\vglue -.15in
\caption{Quantum optical circuit for cubic-phase state preparation. The photon number from the PNR measurement (orange box) conditions  the density matrix that is input to the neural network. Parameters $\alpha_j$, $r_j$, and $\tau_j$ (red boxes) are set by the ML agent at each iteration $j$. $\tau_\text{out}=0$ until the preparation process is over ($\tau_j=0$) which triggers $\tau_\text{out}=1$. The displacement $D(\beta)$ applied prior to the PNR measurement is fixed to $\beta=i2.5$.}
\label{fig:a}
\vglue -.1in
\end{figure}
$B(\tau_j)$ is a variable beamsplitter with transmittivity $\tau_j$ and 
$B(\tau_{\textrm{out}})$ is a switchable mirror with $\tau_{\textrm{out}} = 0$ when
$j<m$ and $\tau_{\textrm{out}} = 1$ when $j=m$. The squeezed state can be realized by an optical parametric oscillator below threshold~\cite{Wu1986} and the loop must be interferometrically stabilized~\cite{Becerra2013}. The PNR detector can be a superconducting transition edge sensor~\cite{Lita2008,Eaton2023} or a superconducting nanowire single-photon detector in the PNR regime~\cite{Cahall2017,Endo2021}. 
Displacements $D(\alpha_j)$ and $D(\beta)$ can be realized by interfering the quantum mode at a highly unbalanced beamsplitter with a weakly transmitted, frequency-degenerate coherent state. 

\subsection{Reinforcement learning concept and prior results}
Reinforcement learning is a branch of ML that seeks to train a learning agent to act optimally in a given environment~\cite{sutton2018reinforcement}. The optimality of an action is informed by a reward signal that the learner receives upon performing the action. The complexity of this learning task arises from the fact that actions taken at one point in time will affect future rewards: for example, selecting an action with a less-than-optimal immediate reward can lead to scenarios where larger rewards are available in the future. Deep reinforcement learning (DRL) is a subfield of reinforcement learning that utilizes deep neural networks as the learning agent and has achieved state of the art performance on many control tasks~\cite{mnih2015human, borah2021measurement, sivak2022model}. 
A lot of the prior work on quantum state preparation has treated the task as an optimization problem in which the parameters of the state preparation circuit are directly optimized to minimize some loss function, e.g.\ the negative of the fidelity of the output state of the circuit with some target state. These methods typically compute the gradients of the circuit parameters with respect to the loss function, using the method of automatic differentiation, and iteratively update the circuit parameters, using the gradient descent algorithm, until a configuration of the circuit parameters is found which minimizes the loss~\cite{goodfellow2016deep}. 
Utilizing this method a measurement-free approach was taken in Ref.~\cite{arrazola2019machine} and Ref.~\cite{kudra2022robust} for the preparation of various non-Gaussian quantum states of interest, the cubic-phase state among them. 
In the case of the former reference the non-Gaussian resources were Kerr nonlinearities whereas in the latter the selective number-dependent arbitrary phase (SNAP) gate supplied the necessary non-Gaussian effects. However, both of these unitaries are challenging to realize in the optical domain. 

Due to this difficulty, postselected PNR-measurement-based optical approaches have also been explored to the same end~\cite{tzitrin2020progress, yao2024riemannian}. 
A major drawback to these approaches is their reliance on post-selection of a specific photon number detection pattern which leads to both low success rates and unfavorable optimization time. The latter stems from the fact that, in the worst case, one must optimize over the set of all possible photon number detection patterns, whose cardinality is on the order of $\mathcal{O}(M^N)$ for an $M$ mode circuit with a photon number cutoff of $N$, to find detection patterns that result in states with both high enough fidelity and probability of success. 

Our approach does not rely on postselection. Rather, it trains the agent to discover states preparation strategies that are adaptive and robust to the inherent randomness of the PNR measurement results. This also comes with the benefit of not needing to examine all possible detection patterns as the agent can learn the underlying distribution of possible measurements outcomes, given a particular configuration of the environment, from past experience. We successfully demonstrated this proposition for the generation of squeezed cat states, with a different optical circuit~\cite{Anteneh2024}.  

\subsection{Reinforcement learning framework and implementation}
In reinforcement learning the interaction between the agent and the environment is modeled using the Markov decision process (MDP) formalism. An MDP is characterized by a set of states $\mathcal{S}$ which describe all possible states of the environment, a set of actions $\mathcal{A}$ which the agent can select from to interact with the environment, a scalar reward function $R(s)$ which determines the reward for reaching state $s \in \mathcal{S}$ and a transition function $\mathcal{T}(s'|s,a)$ which determines the probability that the environment will transition to state $s'$ after the agent selects action $a \in \mathcal{A}$ when the environment is in state $s$. Lastly to be a valid MDP the decision process must also be Markovian which means that the next state $s'$ that the environment transitions to is solely dependent on the current state $s$ and the action taken by the agent while in that state.

The quantum circuit of \fig a can be viewed as a stochastic environment in which the state of the circuit at timestep $j+1$ only depends on the state of the circuit and the action taken at timestep $j$. The state space $\mathcal{S} \subsetneq \mathbb{R}^{N^2}$ consists of the set of all possible flattened density matrices where $N$ is the Hilbert space truncation. Specifically, we define the state at timestep $j$, $s_j$, as the vector
$s_j=[\textrm{Re}(\rho_j^{u}), \textrm{Im}(\rho_j^{u}), 
\textrm{diag}(\rho_j)]$ where $\rho_j^{u}$ denotes all entries of $\rho_j$ that are above its diagonal. The action space $\mathcal{A} \subseteq \mathbb{R}^3$ is the set of actions the agent can select. We represent actions as a vector $a_j = [\tau_j, r_j, \alpha_j]$ where $\tau_j$ is the beamsplitter transmittivity, $r_j$ is the squeezing parameter of the input squeezed vacuum state and $\alpha_j$ is the magnitude of the displacement applied to the input density matrix $\rho_j$. 
In our case the transition function probabilities, $\mathcal{T}(s_{j+1}|s_j,a_j)$, are given by the diagonal elements of the reduced density matrix of the measured mode prior to the PNR measurement.

We defined the reward function $R(s_j)$ at step $j$ as
\begin{equation}
    R(s_j) = \textrm{Tr}^{\frac{\lambda}{10}}(\rho_j)\mathcal{F}^\lambda(\rho_j, \rho_\gamma),
\end{equation}
where $\rho_\gamma$ is the target density matrix, $\mathcal{F}(\rho_j, \rho_\gamma) = \textrm{Tr}(\rho_j\rho_\gamma)$ is the pure-state fidelity, and where we set $\lambda$=55 to tune the penalty on states with fidelities below unity~\cite{porotti2022deep}. The trace of density matrix $\rho_j$ was included in $R(s_j)$ to steer the simulation clear of unphysical results stemming from computation errors caused by Hilbert space truncation, here at 31 photons. The target state
$\rho_\gamma = \ket{\gamma,r,\alpha}\bra{\gamma,r,\alpha}$ was a realistic version of \eq{idealcubic} with an added displacement:
\begin{equation}
    \ket{\gamma,r,\alpha} = D(\alpha)e^{i\gamma Q^3} S(r)\ket{0},
\end{equation}
$\ket{0}$ being the vacuum state, 
$S(r)$=$\exp[\tfrac{r}{2}({a^\dagger}^2-a^2)]$ the squeezing operator of parameter $r$, and 
$D(\alpha)$=$\exp(\alpha a^\dagger - \alpha^* a)$ the displacement operator by $\alpha$. The displacement was intended to maximize the state's occupation probability ($>0.99$) in the 31-photon finite Hilbert space, as per Ref.~\cite{kudra2022robust}. The target state's Wigner function and photon number distribution are shown in \fig b.
\begin{figure}[ht]
\centering
\includegraphics[scale=0.35]{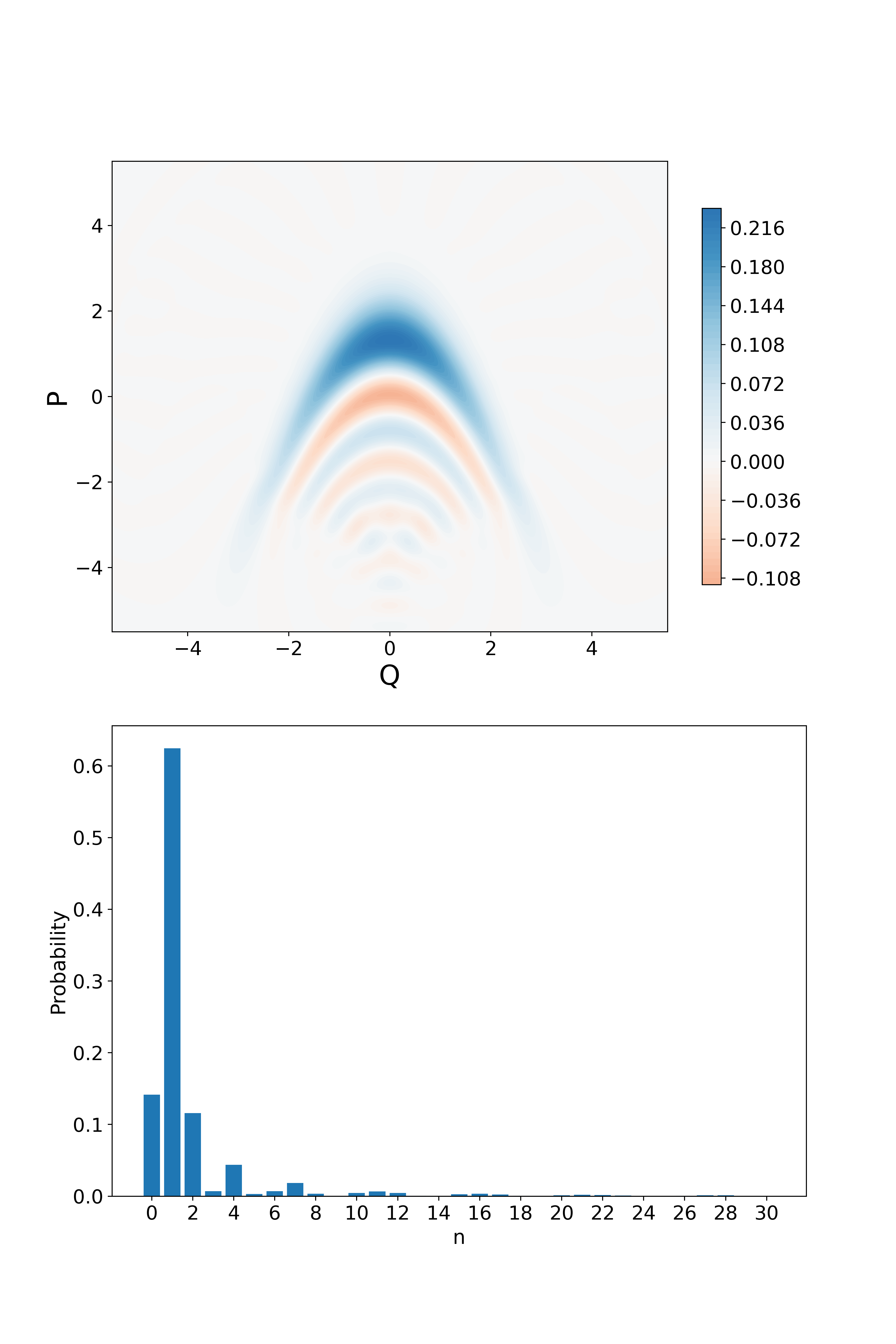}   
\caption{Wigner function and photon number distribution of target cubic-phase state. ($\gamma$,$r$,$\alpha$)=($-0.2$,$-0.7$,$i1.25$).}
\label{fig:b}
\end{figure}
We then proceeded to model the interaction between the agent and the quantum circuit as a MDP. %Moreover, s
Since each episode consisted of a fixed number $m$ of time steps, this constituted a finite-horizon MDP~\cite{sutton2018reinforcement}.

Our reinforcement learning agent  
used proximal policy optimization (PPO)~\cite{schulman2017proximal}. We selected this algorithm due to its on-policy nature, higher robustness to hyperparameter choice compared to other reinforcement learning algorithms~\cite{schulman2017proximal}, and successful use in previous quantum control tasks~\cite{sivak2022model,borah2021measurement}. The on-policy attribute is particularly important for reducing the risk of training instability in our implementation of PPO, which utilizes both deep neural networks and bootstrapping. This is known as the `deadly triad,' which refers to the possibly detrimental, joint action of off-policy learning, universal function approximation, and bootstrapping~\cite{sutton2018reinforcement}. 
We used the Python libraries \texttt{StrawberryFields}~\cite{killoran2019strawberry} and \texttt{StableBaselines3}~\cite{raffin2021stable} to simulate the quantum circuit and implement the PPO algorithm respectively. The hyperparameters used for the PPO algorithm are shown in Table \ref{table:1}. 
\begin{table}
\centering
\begin{tabular}{ |c|c|c| } 
 \hline
 Parameter & Value\\ 
 \hline
 \texttt{gamma} & 0.999\\ 
 \hline
 \texttt{n\_steps} & 35,000\\
 \hline
 \texttt{batch\_size} & 5,000\\ 
 \hline
 \texttt{n\_epochs} & 15\\
 \hline
 \texttt{clip\_range} & 0.2\\
 \hline
 \texttt{optimizer} & Adam\\
 \hline
 \texttt{learning\_rate} & 0.001\\
 \hline
 \texttt{max\_grad\_norm} & 0.5\\
 \hline
 \texttt{vf\_coef} & 0.5\\
 \hline 
 \texttt{ent\_coef} & 0.0\\
 \hline
 \texttt{activation\_fn} & $\tanh$\\
 \hline
\end{tabular}
\caption{Hyperparameters used for the PPO algorithm during training.}
\label{table:1}
\end{table}
We tuned the hyperparameters \texttt{gamma}, \texttt{batch\_size} and \texttt{learning\_rate} via a grid search over the values $[0.99, 0.999]$, $[1000,2500,5000]$ and $[10^{-2}, 10^{-3}, 10^{-4}]$ respectively. 
The \texttt{StableBaselines3} implementation of PPO utilizes an actor-critic framework, so our agent was composed of two deep neural networks: the actor network, which performs action selection for a given state $s$, and the critic network, which is meant to learn the value $V(s)$ of state $s$. Both networks have the same architecture: three hidden layers of sizes 256, 128, and 64, respectively, with $\tanh$ as the non-linear activation function.
The function of the hyperparameters listed in Table \ref{table:1}, a summary of PPO, and a more detailed introduction to the relevant reinforcement learning concepts can be found in the supplemental material of Ref.~\cite{Anteneh2024}.
To speed up the training process we ran 40 environments in parallel for faster training data collection. The initial input state $\rho_0$ for each episode was a 10 dB squeezed state ($r$=$1.15$). The maximum magnitude of squeezing and displacement the agent can apply at each step were set to $|r_\text{max}|=1.15$ and $|\alpha_\text{max}|=2.25$ respectively. We fixed the displacement before the PNR measurement to $\beta=i2.5$.

\section{Simulation results for cubic-phase state generation}
\subsubsection{Lossless Case}
The agent was trained over 5.7 million time steps with episodes of length $m=50$. \Fig c shows the fidelity of the state at the end of each episode with the target state over the course of training. 
\begin{figure}[htb]
\centering
\includegraphics[width=\columnwidth]{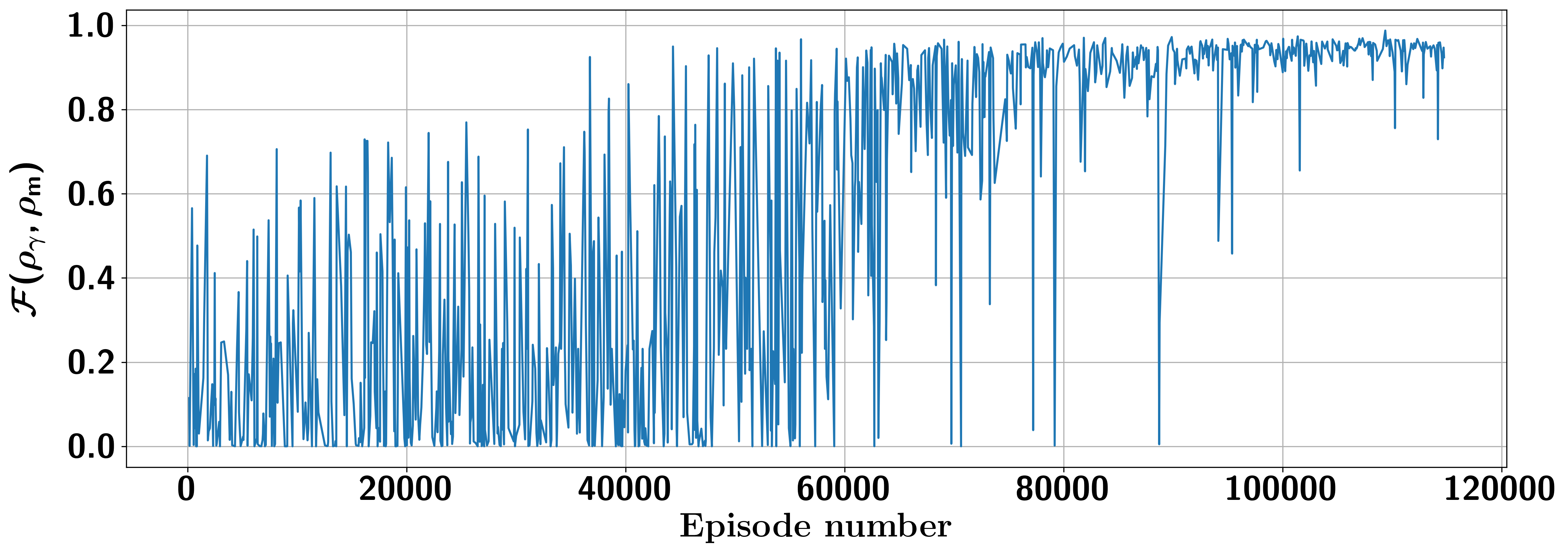}   
\caption{Terminal state, $\rho_m$, fidelity with the target state, $\rho_\gamma$, per episode during lossless training}
\label{fig:c}
\end{figure}
The plot indicates that the agent has converged to a policy that maximizes the fidelity of the final state for each episode. Due to the stochastic nature of the environment there still remain some episodes which terminate with a state that has less than optimal fidelity. Once trained we evaluated the agents performance by running five sets of 200 evaluation episodes. Each set of episodes utilized a different random-number seed for simulating PNR detection. 

The results from the evaluation are collected in \fig d. \begin{figure*}[ht]
\centering
\includegraphics[width=\linewidth, scale=0.175]{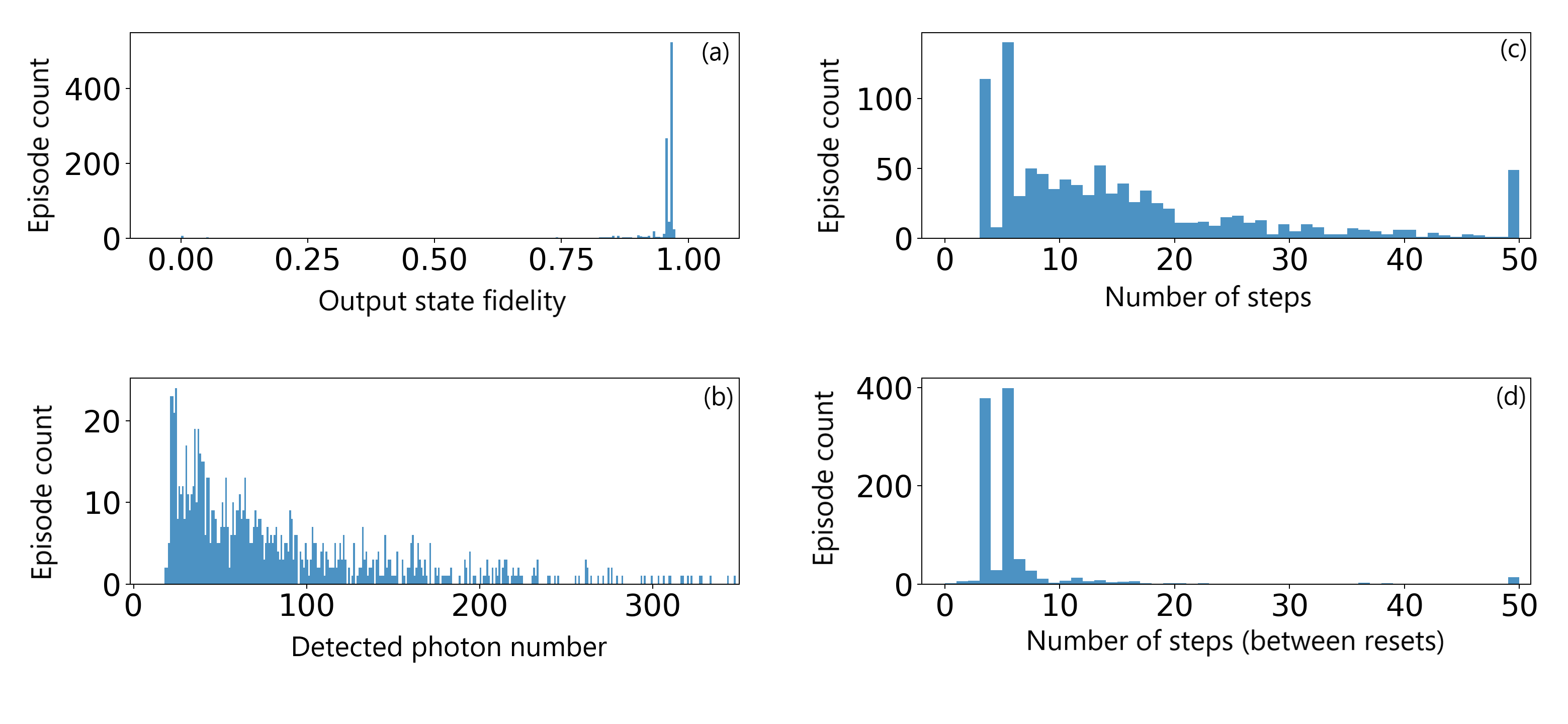}   
\caption{Histograms for 1000 lossless generation episodes ($m$ = 50): (a) output state fidelity; (b) total number of photons detected per episode before $\tau_j=0$; (c)
steps per episode, with resets ($\tau_j=1$), before $\tau_j=0$; (d) steps per episode, between resets  ($\tau_j=1$), before $\tau_j=0$.}
\label{fig:d}
\end{figure*}
We observed interesting emergent behaviors from the trained neural network: \\
{\it(i)} {\it After} reaching high fidelities, \fig da, the agent set the beamsplitter's transmittivity $\tau_j$ to $0$ and didn't change it for the remaining  of the episode. Instances of $\tau_j$=$0$ at low fidelity didn't remain unchanged.\\
{\it(ii)} After reaching high fidelities and setting $\tau_j$ to $0$, the agent  frequently selected non-zero values of $\alpha_j$ for a few more steps before eventually setting $\alpha_j$ to $0$ as well for the remaining time steps of the episode. We interpreted this behavior as the agent recognizing that the current input state to the loop, $\rho_j$, was the target cubic-phase state up to corrective displacements which it  applied after setting the transmittivity to $0$. Note that these ``final'' displacements sometimes exhibited oscillatory behaviors in the lossy case (see below).\\
{\it(iii)} The agent occasionally set $\tau_j$ to $1$, i.e.,  `reset' the environment to restart from the initial squeezed state input. This indicates that the agent  learned to identify when a particular input state $\rho_j$ is unlikely to be transformed into the target state with further iterations of the loop and thus found it more beneficial to simply reset the circuit and try again. From \fig dd we can see that in the vast majority of cases no more than ten iterations of the loop were needed before the agent generated a state it was satisfied with. \\
{\it(iv)} The agent could also set $\tau_j=0$ at very low fidelity but not stay at this value in the subsequent steps. These corresponded to some of the failure episodes.\\
{\it(v)} About 50 episodes out of 1000 reached the maximum step number of 50, \fig dc. Out of these, 19 were failures, the rest were  successful events in which the agent would keep ``adjusting'' the state with tiny, inconsequential displacements. We believe that more training and hyperparameter tuning should help these cases converge faster. \\
{\it(vi)} We counted only 5 instances out of 1000 where the agent appeared ``lost,'' i.e., stuck at low fidelity at every step with no observable improvement. 

We conducted a detailed visual examination of all the resulting Wigner functions from all 1000 episodes (thus crudely assessing  minimum negativity), as well as checked the final fidelities, maximum step number, and that ${\rm Tr}[\rho_j]\geqslant0.98$ at all steps after the last reset. We found that a {\it bona fide} cubic-phase state was generated 96\% of the time, with some variations of $\alpha$ and $\gamma$ with respect to the target state, sometimes resulting in lower fidelities, \fig da. These variations were not judged detrimental for two reasons: first, they constituted {\it bona fide} cubic-phase states; second, all output states can be uniquely tagged by their record of measurement and parameter values, which enables the creation of lookup tables that can uniquely identify the generated state at every turn. Because of all this, all outputs are usable. Note finally that concatenating two cubic gates of parameters $\gamma_1$ and $\gamma_2$ generates, in principle, a gate of parameter $\gamma_1 + \gamma_2$ and storing cubic-phase states in quantum memories (or delay lines) can allow ``breeding'' larger ones. 

\subsubsection{Lossy Case}

We also trained an agent to prepare the same target state when the circuit utilized a PNR detector with only 99\% quantum efficiency.
Due to the noise introduced into the environment by the lossy detector the agent was trained for a longer period of time than in the lossless case: approximately 6 million timesteps with the episode length remaining unchanged at $m=50$. The same number of evaluation episodes was simulated.
Surprisingly, upon closer examination of each episodes final Wigner functions we found that on average the resulting Wigner function was that of a cubic-phase state with little degradation in quality compared to the lossless case. 

However the addition of loss did result in a marked change in the agents behavior during evaluation. The agent would still set $\tau_j$ to 0 upon arriving at a state with high fidelity but for the remainder of the episode the agent would repeatedly apply displacements of alternating sign. This resulted in the fidelity oscillating between near $0$ and the initial high fidelity that caused the agent to set the transmittivity to 0. This oscillation in fidelity was also observed during training as can be seen in \fig e, 
\begin{figure}[ht]
\centering
\includegraphics[width=\columnwidth]{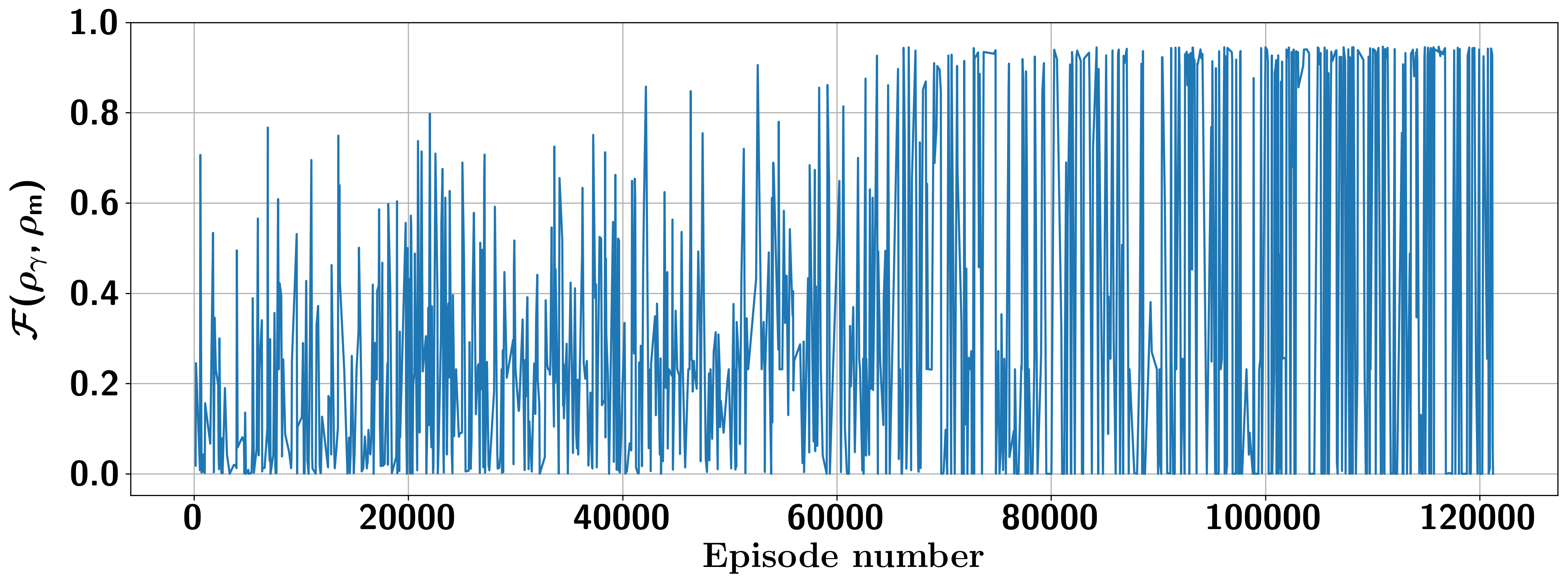}   
\caption{Terminal state, $\rho_m$, fidelity with the target state, $\rho_\gamma$, per episode during lossy training}
\label{fig:e}
\end{figure}
which shows the fidelity of the
terminal state with the target state over the course of training with the lossy detector. When the detector efficiency was further lowered to 90\% we found the agent was unable to learn a policy that resulted in any cubic-phase state being generated.  

\section{Quartic-phase algorithm}

It is known that the quartic-phase gate $\exp(i\delta Q^4)$ of parameter $\delta$ can be realized by assembling 29 gates, including 15 cubic-phase gates~\cite{Kalajdzievski2019}. One question is whether the methods presented above wouldn't be applicable to the {\it direct} generation of a quartic-phase state. Our full numerical ML simulations of this case are still a work in progress as they require considerable run-time and error optimizations compared to the cubic-phase case. (One main reason for this is the necessity to raise the photon number cutoff from 31 to at least 60, if not higher.)  However, we have identified a fundamental, though postselected, quantum optical algorithm that can form the basis of subsequent, deterministic ML implementation. We present this algorithm here. 

\subsection{Quantum algorithm}

\begin{figure}[ht]
\centering
\begin{subfigure}{\columnwidth}
\includegraphics[width=\columnwidth]{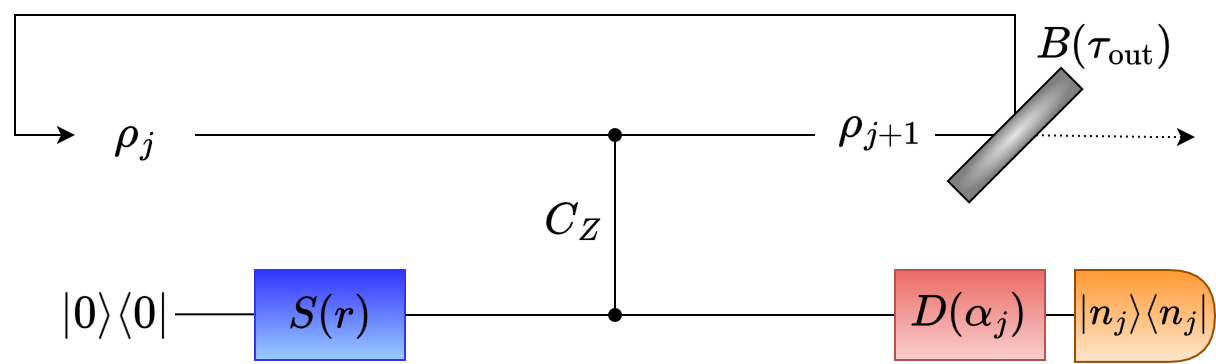}   
\caption{Quantum circuit for quartic state preparation. $\rho_0=S(r)\ket0\bra0S^\dag(r)$.}
\label{fig:cla}
\end{subfigure}
\begin{subfigure}{\columnwidth}
\includegraphics[width=\columnwidth]{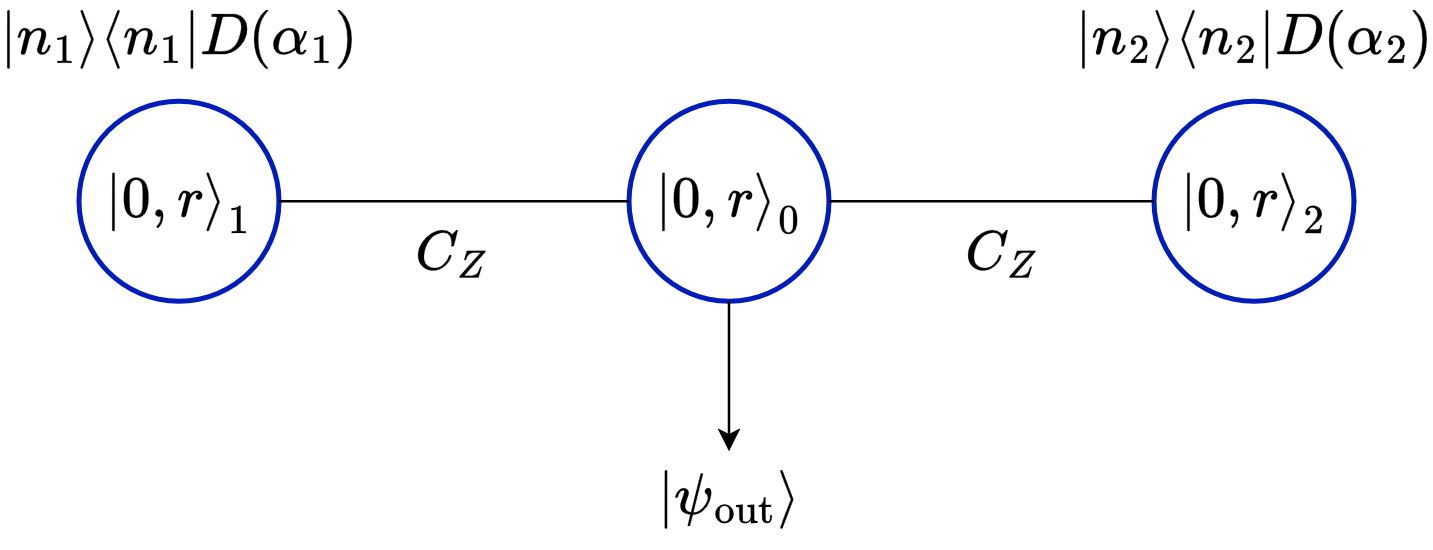}   
\caption{CV Cluster state~\cite{Zhang2006,Gu2009,Pfister2019} and measurements for quartic state preparation. The $C_Z$ operator is $C_{Zjk}=\exp(iQ_jQ_k)$.} 
\label{fig:clb}
\end{subfigure}
\caption{Circuit-based diagram and equivalent measurement-based diagram of the quartic-phase state preparation algorithm.}
\label{fig:cl}
\end{figure}

The intuition is to ``stamp'' the Wigner function with a displaced Fock state at nearly opposite azimuths in phase space, so as to generate the required cubic-polynomial contour of a quartic-phase state. This can be effected by the circuit presented in \fig{cla}, also presented in its MBQC version in \fig{clb}. 
Here, a big difference is that the squeezing is held constant, the only adjustable parameters being the phase of the second displacement before the PNR. Posing the PNR displacements $\alpha_{1,2}=|\alpha_{1,2}|\exp(i\phi_{1,2})$, we took $|\alpha_{1,2}|= 4.5$, $\phi_1=\pi$, and $\phi_2\simeq 0$ a nonzero adjustable parameter. ($\phi_2=0$ is too symmetric and produces zero quarticity.) The constant squeezing in \fig{cl} was set to $r=1.38$ (12 dB). 

\subsection{Postselected preliminary results}
\begin{figure*}[ht]
\centering
\includegraphics[width=1\textwidth]{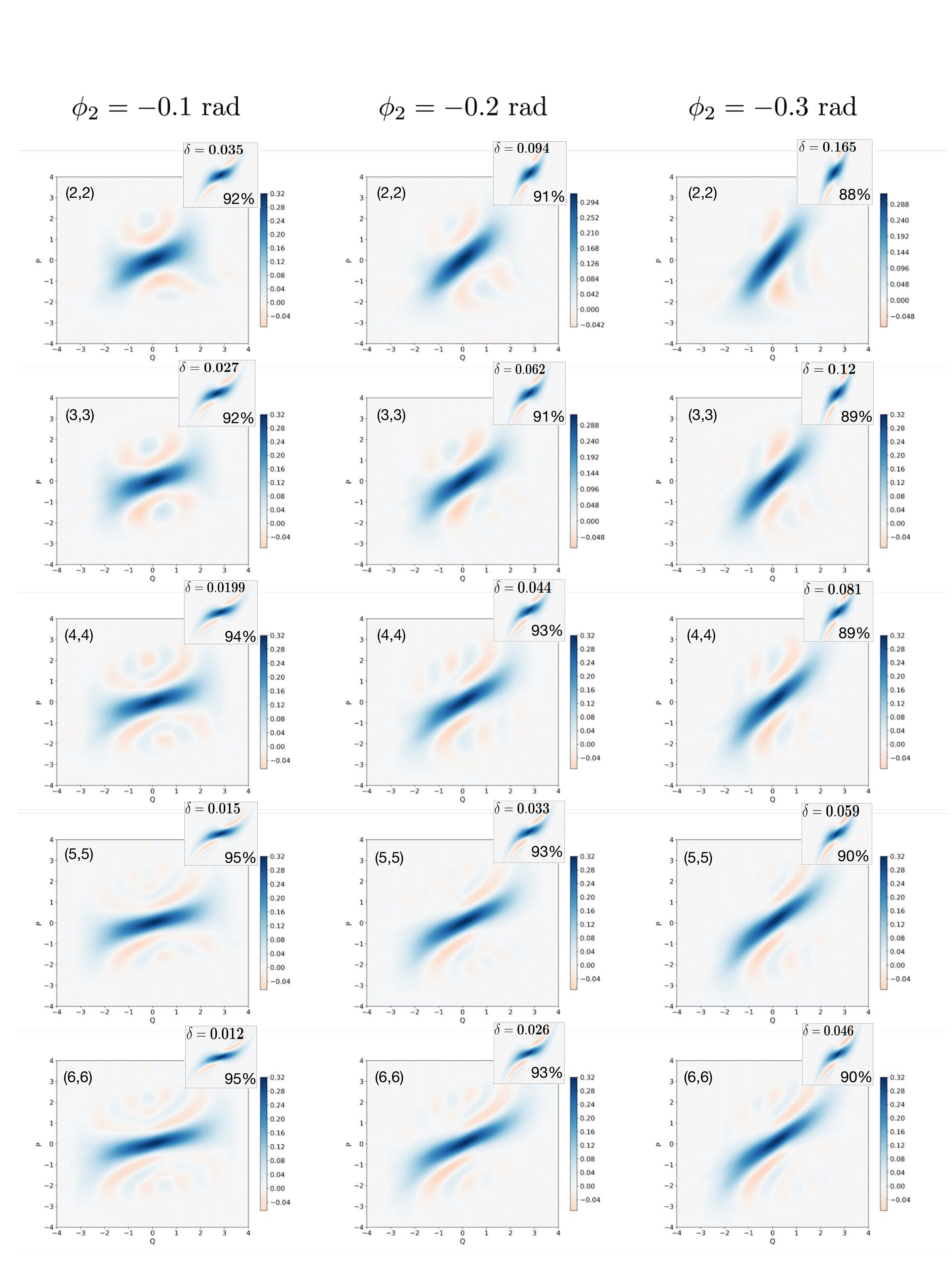}   
\caption{Resulting Wigner functions of two rounds of PNR detection on the cluster state in \fig{cl}, for different values of $\phi_2$. In each case, the PNR results ($n_1$,$n_2$) are indicated in the top left corner and the top right inset depicts the best-fit Wigner function with its fidelity and quarticity.}
\label{fig:res}
\end{figure*}

We define the target quartic-phase state as
\begin{equation}
    \ket{\delta,s}= e^{i\delta Q^4} S(s)\ket{0},
\end{equation}
where $\delta$ is the ``quarticity'' and $s$ is a squeezing parameter. To determine which quartic-phase state was generated by our two-step circuit we used the Nelder-Mead optimization algorithm to find the values of $\delta$ and $s$ that maximized the fidelity between $\ket{\delta,s}$ and the output state from our circuit for each set of postselected photon number measurements. For the simulations in this section we used a considerably higher Hilbert space truncation of 60 photons. The best results obtained for output states and their corresponding optimized quartic-phase states are shown in \fig{res} for $\phi_2=-0.1,-0.2,-0.3$ rad.
In an unsurprising confirmation of our original intuition, the best results were obtained for $n_1=n_2$, which is the reason why this algorithm requires postselection. One of the outcomes expected from our upcoming ML extension of this algorithm will clearly be the ability to cope with $n_1\neq n_2$. The gist of the results in \fig{res} is that engineering quantum interference in phase space~\cite{Schleich} is a tenet of quantum state generation, here turning circular Fock state contours into the characteristic cubic-polynomial ripples of a quartic-phase state. This seemed to happen more effectively at higher photon numbers, where the fidelity with a genuine quartic-phase state was reasonably high, which validated the algorithm. We did verify that this fidelity was always larger than the fidelity to a displaced squeezed state. The value of $\phi_2$ seemed to play a role as well and its possible dependence on $n_1-n_2$ is another landscape that we look forward to ML exploring.  

\section{Conclusion and prospective work}
We have proposed a method based on deep reinforcement learning for the near-deterministic preparation of cubic-phase states. Besides being QC-efficient, this method requires less squeezing and photon number detection capability than proposals so far, and can be implemented with components that are readily available in current experimental setups, the only non-Gaussian resource being PNR measurements~\cite{Eaton2023}. We also gave an algorithm for the direct, although probabilistic, generation of quartic-phase gates.  The extension of this algorithm to the near-deterministic generation of quartic-phase states, using DRL simulations, is a work in progress. These simulations require a much larger Hilbert space which entails a significant growth of needed computational resources.

\textit{Acknowledgments}---% 
We acknowledge support from NSF grants PHY-2112867 and ECCS-2219760. We  thank U. of Virginia Research Computing for providing access to the Rivanna computing cluster. 

%\section{Competing Interests}  
%The authors declare no conflicts of interest.
%\section{Data availability} 
%Data pertaining to this paper's results may be obtained from the authors upon reasonable request.

%\section{Code availability} 
%Computer code used to generate the data presented may be obtained from the authors upon reasonable request.

\bibliography{Pfister, Anteneh}

\end{document}